\documentclass[11pt]{cernrep}
\usepackage{graphicx}
\usepackage{here}

\def\bea{\begin{eqnarray}}
\def\eea{\end{eqnarray}}
\def\be{\begin{equation}}
\def\ee{\end{equation}}
\newcommand{\ub}[1]{\underline{#1}}
\def\del{\partial}
\def\g{\gamma}
\def\ha{\frac{1}{2}}
\def\psibar{\overline{\psi}}

\begin{document}

\title{%
%
%
\begin{flushright} {\rm \normalsize UMN-D-03-6} \end{flushright}
Nonperturbative light-front methods%
%
%
\footnote{\ To appear in the proceedings of 
the International Light-Cone Workshop: Hadrons and Beyond,
the Institute for Particle Physics Phenomenology, Durham, UK,
August 5-9, 2003.}%
}
\author{J.R. Hiller}
\institute{Department of Physics, University of Minnesota-Duluth, 
Duluth, Minnesota~~55812, USA}
\maketitle
\begin{abstract}
Two methods for the nonperturbative solution of field-theoretic
bound-state problems, based on light-front coordinates, are
briefly reviewed.  One uses Pauli--Villars regularization
and the other supersymmetry.  Applications to Yukawa theory
and super Yang--Mills theory with fundamental matter are 
emphasized.
\end{abstract}

\section{INTRODUCTION}

There have been a number of calculations~\cite{DLCQreview}
using light-front coordinates~\cite{Dirac} as a convenient
means of attacking field-theoretic problems nonperturbatively,
particularly in 1+1 dimensions.  Efforts in more dimensions
have generally been less successful, however, due to the
need for regularization and renormalization.  Here two
approaches that include consistent regularization are
briefly reviewed in the context of specific model 
calculations~\cite{TwoParticles,SQCD-CS}.  One approach
is Pauli--Villars (PV) regularization~\cite{PauliVillars},
where massive negative-metric particles are added to a 
theory to provide the necessary cancellations~\cite{bhm,Special}.
The other is supersymmetry~\cite{SDLCQreview}.  These are
not the only approaches available on the light front; in particular, 
one can find the transverse lattice technique~\cite{TransLattice}
and similarity transformations~\cite{Glazek} discussed
elsewhere in this volume.

The primary numerical method is discrete light-cone 
quantization (DLCQ)~\cite{PauliBrodsky}, in which one 
imposes a discrete momentum
grid, with length scales $L$ and $L_\perp$, as
$p^+\rightarrow \pi n/L$ and 
${\mathbf p}_\perp\rightarrow\pi {\mathbf n}_\perp/L_\perp$,
and approximates integrals in the mass-squared eigenvalue problem
with trapezoidal sums.  The continuum limit $L\rightarrow\infty$
can be exchanged for a limit in terms of the integer resolution
$K\equiv L P^+/\pi$, because light-cone momentum fractions
$x_i\equiv p_i^+/P^+$ are measured in units of $1/K$.

For supersymmetric theories there is a supersymmetric version
of DLCQ (SDLCQ) that preserves
supersymmetry within the discrete approximation~\cite{Sakai}.
This is accomplished by discretizing the supercharge $Q^-$
and constructing the Hamiltonian $P^-$ from the superalgebra
via the anticommutator: $P^-=\{Q^-,Q^-\}/2\sqrt{2}$.  This
$P^-$ and the DLCQ $P^-$ are equivalent in the $K\rightarrow\infty$
limit.

The matrix eigenvalue problems that result from these
discretizations are large but sparse.  An efficient
means for extracting a few lowest eigenvalues and their
eigenvectors is the Lanczos algorithm~\cite{Lanczos}.
In the case of PV-regulated theories, with their indefinite
metrics, a special form~\cite{Special} based on the 
biorthogonal algorithm~\cite{biorthogonal} is required.  In
either case, the process is an iterative one that
generates a tridiagonal matrix of much smaller size,
which is easily diagonalized to yield approximate
eigenvalues and eigenvectors.  The number of iterations
and the size of the tridiagonal matrix are determined
by the rate of convergence.  The primary difficulty that 
arises is that round-off error causes spurious copies to
appear in the derived spectrum; however, there are techniques
for removing them~\cite{Cullum}.

In the remaining sections we discuss an application of
PV regularization to Yukawa theory and a study of
supersymmetric QCD (SQCD) with a Chern--Simons (CS)
term in the large-$N_c$ approximation.

\section{YUKAWA THEORY}

As the action of the PV-regulated Yukawa theory, we take
\bea
\lefteqn{S=\int d^4x
\left[\ha(\del_\mu\phi_0)^2-\ha\mu_0^2\phi_0^2
-\ha(\del_\mu\phi_1)^2+\ha\mu_1^2\phi_1^2
 +\frac{i}{2}\left(\psibar_0\g^\mu\del_\mu-(\del_\mu\psibar_0)\g^\mu\right)
     \psi_0\right.}  \\
&& \left.
  -m_0\psibar_0\psi_0 
  -\frac{i}{2}\left(\psibar_1\g^\mu\del_\mu-(\del_\mu\psibar_1)\g^\mu\right)
      \psi_1+m_1\psibar_1\psi_1
-g(\phi_0 + \phi_1)(\psibar_0 + \psibar_1)(\psi_0 + \psi_1)\right],
\nonumber
\eea
with the subscript 0 indicating physical fields and 1 indicating PV fields.
When fermion pairs are excluded, this action provides a light-cone
Hamiltonian of the form
\bea \label{eq:YukawaP-}
\lefteqn{P^-=
   \sum_{i,s}\int d\underline{p}
      \frac{m_i^2+{\mathbf p}_\perp^2}{p^+}(-1)^i
          b_{i,s}^\dagger(\underline{p}) b_{i,s}(\underline{p})
      +\sum_{j}\int d\underline{q}
          \frac{\mu_j^2+{\mathbf q}_\perp^2}{q^+}(-1)^j
              a_j^\dagger(\underline{q}) a_j(\underline{q}) } \\
   && +\sum_{i,j,k,s}\int d\underline{p} d\underline{q}\left\{
      \left[ V_{-2s}^*(\underline{p},\underline{q})
             +V_{2s}(\underline{p}+\underline{q},\underline{q})\right]
                 b_{j,s}^\dagger(\underline{p})
                  a_k^\dagger(\underline{q})
                   b_{i,-s}(\underline{p}+\underline{q})\right.  \nonumber \\
      &&\left.\rule{0.5in}{0in}
           +\left[U_j(\underline{p},\underline{q})
                    +U_i(\underline{p}+\underline{q},\underline{q})\right]
               b_{j,s}^\dagger(\underline{p})
                a_k^\dagger(\underline{q})b_{i,s}(\underline{p}+\underline{q})
                    + h.c.\right\},  \nonumber
\eea
where $a^\dagger$ creates a boson and $b^\dagger$ a fermion,
\be
U_j(\underline{p},\underline{q})
   \equiv \frac{g}{\sqrt{16\pi^3}}\frac{m_j}{p^+\sqrt{q^+}},\;\;
V_{2s}(\underline{p},\underline{q})
   \equiv \frac{g}{\sqrt{8\pi^3}}
   \frac{\mathbf{\epsilon}_{2s}^{\,*}\cdot\mathbf{p}_\perp}{p^+\sqrt{q^+}},\;\;
{\mathbf \epsilon}_{2s}\equiv-\frac{1}{\sqrt{2}}(2s,i), 
\ee
and
\be
\left[a_i(\underline{q}),a_j^\dagger(\underline{q}')\right]
          =(-1)^i\delta_{ij}
            \delta(\underline{q}-\underline{q}'), \;\;
\left\{b_{i,s}(\underline{p}),b_{j,s'}^\dagger(\underline{p}')\right\}
     =(-1)^i\delta_{ij}   \delta_{s,s'}
            \delta(\underline{p}-\underline{p}').
\ee
The eigenfunction for the dressed-fermion state is expanded
in a Fock basis as
\bea
\lefteqn{\Phi_+(\ub{P})=\sum_i z_i b_{i+}^\dagger(\ub{P})|0\rangle
  +\sum_{ijs}\int d\ub{q} f_{ijs}(\ub{q})b_{is}^\dagger(\ub{P}-\ub{q})
                                       a_j^\dagger(\ub{q})|0\rangle}&& \\
 && +\sum_{ijks}\int d\ub{q}_1 d\ub{q}_2 f_{ijks}(\ub{q}_1,\ub{q}_2)
       \frac{1}{\sqrt{1+\delta_{jk}}}   b_{is}^\dagger(\ub{P}-\ub{q}_1-\ub{q}_2)
                 a_j^\dagger(\ub{q}_1)a_k^\dagger(\ub{q}_2)|0\rangle 
 +\ldots       \nonumber
\eea
It is normalized according to
$\Phi_\sigma^{\prime\dagger}\cdot\Phi_\sigma
=\delta(\underline{P}'-\underline{P})$.
The wave functions satisfy a coupled system of equations, derived
from the fundamental mass-squared eigenvalue problem $P^+P^-\Phi_+=M^2\Phi_+$
to be
\bea
m_i^2z_i&+& \sum_{i',j}(-1)^{i'+j} P^+ \int^{P^+} d\ub{q}
  \left\{ f_{i'j-}(\ub{q})[V_+(\ub{P}-\ub{q},\ub{q})+V_-^*(\ub{P},\ub{q})]
  \right. \\
  &&\left.
   + f_{i'j+}(\ub{q})[U_{i'}(\ub{P}-\ub{q},\ub{q})+U_i(\ub{P},\ub{q})]\right\}
   = M^2z_i,\nonumber
\eea
\bea \label{eq:oneboson}
\lefteqn{
\left[\frac{m_i^2+q_\perp^2}{1-y}+\frac{\mu_j^2+q_\perp^2}{y}\right]
  f_{ijs}(\ub{q}) 
  +\sum_{i'}(-1)^{i'}\left\{
    z_{i'}\delta_{s,-}[V_+^*(\ub{P}-\ub{q},\ub{q})+V_-(\ub{P},\ub{q})] 
                               \right.}\\
  && \rule{2in}{0in}\left.
    +z_{i'}\delta_{s,+}[U_i(\ub{P}-\ub{q},\ub{q})+U_{i'}(\ub{P},\ub{q})]\right\}
    \nonumber \\
    &&+2\sum_{i',k}\frac{(-1)^{i'+k}}{\sqrt{1+\delta_{jk}}}P^+ \int^{P^+-q^+}
    d\ub{q}' \left\{f_{i'jk,-s}(\ub{q},\ub{q}')
       [V_{2s}(\ub{P}-\ub{q}-\ub{q}',\ub{q}')+V_{-2s}^*(\ub{P}-\ub{q},\ub{q}')]
       \right. \nonumber \\
           &&\left.+f_{i'jks}(\ub{q},\ub{q}')
             [U_{i'}(\ub{P}-\ub{q}-\ub{q}',\ub{q}')+U_i(\ub{P}-\ub{q},\ub{q}')]
             \right\} = M^2f_{ijs}(\ub{q}),\nonumber
\eea
\bea
\lefteqn{\left[\frac{m_i^2+(\mathbf{q}_{1\perp}+\mathbf{q}_{2\perp})^2}{1-y_1-y_2}
    +\frac{\mu_j^2+q_{1\perp}^2}{y_1}+\frac{\mu_k^2+q_{2\perp}^2}{y_2}\right]
        f_{ijks}(\ub{q_1},\ub{q_2})} \\
    &&+\sum_{i'}(-1)^{i'}\frac{\sqrt{1+\delta_{jk}}}{2}P^+ 
        \left\{f_{i'j,-s}(\ub{q_1})
[V_{-2s}^*(\ub{P}-\ub{q_1}-\ub{q_2},\ub{q_2})+V_{2s}(\ub{P}-\ub{q_1},\ub{q_2})]
 \right. \nonumber \\
 &&+ f_{i'js}(\ub{q_1})
[U_i(\ub{P}-\ub{q_1}-\ub{q_2},\ub{q_2})+U_{i'}(\ub{P}-\ub{q_1},\ub{q_2})]
 \nonumber \\
 &&  + f_{i'k,-s}(\ub{q_2})
[V_{-2s}^*(\ub{P}-\ub{q_1}-\ub{q_2},\ub{q_1})+V_{2s}(\ub{P}-\ub{q_2},\ub{q_1})]
  \nonumber \\
 &&  \left. + f_{i'ks}(\ub{q_2})
[U_i(\ub{P}-\ub{q_1}-\ub{q_2},\ub{q_1})+U_{i'}(\ub{P}-\ub{q_2},\ub{q_1})]
\right\}+\ldots  = M^2f_{ijks}(\ub{q_1},\ub{q_2}). \nonumber
\eea
These equations can be approximated directly by DLCQ.

We now have a well-defined numerical problem.  The PV particles
are kept in the DLCQ basis and provide the necessary counterterms.
The range of the now-finite transverse integrations is cut off by imposing
$p_{i\perp}^2/x_i<\Lambda^2$ for each particle in a Fock state,
to reduce the matrix problem to a finite size.  The transverse 
momentum indices $n_x$ and $n_y$ are limited by the transverse resolution
$N$.  The bare parameters
$g$ and $m_0$ are fixed by fitting ``physical'' constraints, such 
as specifying the dressed-fermion mass $M$ and its radius.  The
limits of infinite resolution, infinite (momentum) volume, and
infinite PV masses can then be explored.

This process can be studied analytically in the case of
a one-boson truncation~\cite{TwoParticles}.
The one-boson wave functions are
\bea
f_{ij+}(\underline{q})&=&
   \frac{P^+}{M^2-\frac{m_i^2+q_\perp^2}{1-q^+/P^+}
                -\frac{\mu_j^2+q_\perp^2}{q^+/P^+}}
\left[(\sum_k (-1)^{k+1}z_k)U_i(\underline{P}-\underline{q},\underline{q})
      +\sum_k (-1)^{k+1}z_kU_k(\underline{P},\underline{q})\right],
 \nonumber \\
f_{ij-}(\underline{q})&=&
   \frac{P^+}{M^2-\frac{m_i^2+q_\perp^2}{1-q^+/P^+}
                -\frac{\mu_j^2+q_\perp^2}{q^+/P^+}}
(\sum_k (-1)^{k+1}z_k)V_+^*(\underline{P}-\underline{q},\underline{q}).
\eea
The bare-fermion amplitudes and the coupling satisfy a pair of
algebraic equations
\bea \label{eq:onefermion}
(M^2-m_i^2)z_i &=&
 g^2\mu_0^2 (z_0-z_1)J+g^2 m_i(z_0m_0-z_1m_1) I_0
\nonumber \\
  &&+g^2\mu_0[(z_0-z_1)m_i+z_0m_0-z_1m_1] I_1,
\eea
with
\bea
I_n&=&\int\frac{dy dq_\perp^2}{16\pi^2}
   \sum_{jk}\frac{(-1)^{j+k}}{M^2-\frac{m_j^2+q_\perp^2}{1-y}
                                   -\frac{\mu_k^2+q_\perp^2}{y}}
   \frac{(m_j/\mu_0)^n}{y(1-y)^n}, \\
J&=&\int\frac{dy dq_\perp^2}{16\pi^2}
   \sum_{jk}\frac{(-1)^{j+k}}{M^2-\frac{m_j^2+q_\perp^2}{1-y}
                                   -\frac{\mu_k^2+q_\perp^2}{y}}
   \frac{(m_j^2+q_\perp^2)/\mu_0^2}{y(1-y)^2}=\frac{M^2}{\mu_0^2}I_0.
\eea
The solution is
\be \label{eq:gofm}
g^2=-\frac{(M\mp m_0)(M\mp m_1)}{(m_1-m_0)(\mu_0 I_1\pm MI_0)}, \;\;
\frac{z_1}{z_0}=\frac{M \mp m_0}{M \mp m_1}.
\ee
Ref.~\cite{TwoParticles} contains many subsequent results.

For a two-boson truncation, the solution is no longer 
analytic, but the coupled equations can be reduced
to eight equations for the two-particle amplitudes only,
which are of the form
\bea
\lefteqn{\left[M^2
  -\frac{m_i^2+q_\perp^2}{1-y}-\frac{\mu_j^2+q_\perp^2}{y}\right]
f_{ijs}(y,q_\perp)=
\frac{g^2}{16\pi^2}\sum_a\frac{I_{ija}(y,q_\perp)}{1-y}f_{ajs}(y,q_\perp)}
 \\
 &&  +\frac{g^2}{16\pi^2}\sum_{abs'}\int_0^1dy'dq_\perp^{\prime 2}
   J_{ijs,abs'}^{(0)}(y,q_\perp;y',q'_\perp)f_{abs'}(y',q'_\perp) \nonumber \\
   && +\frac{g^2}{16\pi^2}\sum_{abs'}\int_0^{1-y}dy'dq_\perp^{\prime 2}
   J_{ijs,abs'}^{(2)}(y,q_\perp;y',q'_\perp)f_{abs'}(y',q'_\perp),
   \nonumber
\eea
with the angular dependence removed via
$\sqrt{P^+}f_{ij+}(\ub{q})=f_{ij+}(y,q_\perp)$ and
$\sqrt{P^+}f_{ij-}(\ub{q})=f_{ij-}(y,q_\perp)e^{i\phi}$.
Here $I$ is a computable self-energy,
$J^{(0)}$ is the kernel due to bare-fermion intermediate states,
and $J^{(2)}$ is the kernel due to two-boson intermediate states.

\begin{figure}[hpbt]
\begin{center}
\includegraphics[width=13cm]{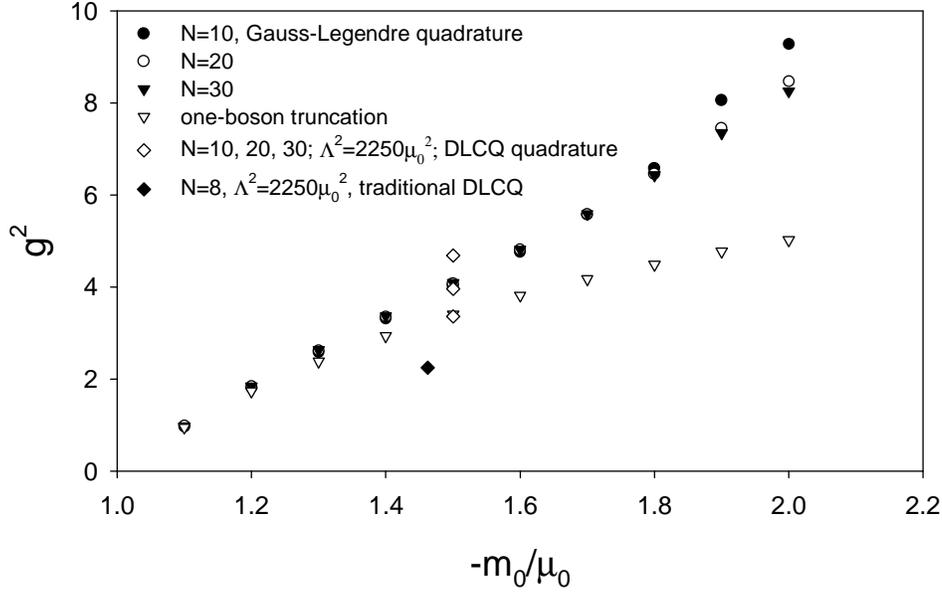}
\caption{The Yukawa coupling as a function of the bare fermion
mass for the two-boson and (exactly soluble) one-boson truncations.
The dressed-fermion mass is $M=\mu_0$, and 
the PV masses are $m_1=\mu_1=15\mu_0$.  The longitudinal resolution
is $K=20$; the transverse resolutions $N$ are specified in the
legend.}
\label{fig:g2}
\end{center}
\end{figure}

Although one could impose a transverse cutoff and discretize these
equations as per DLCQ, alternative quadratures are more efficient. 
In particular, the transverse momentum $q_\perp$ can be mapped to
a finite range that compresses the wave function's tail to a
relatively small region, so that a Gauss--Legendre quadrature can
yield a good approximation.  A comparison of results is given in
Fig.~\ref{fig:g2}.
%

\section{SUPERSYMMETRIC QCD}

As the action for (2+1)-dimensional ${\cal N}=1$ SQCD-CS theory, we consider
\bea\label{action}
S&=&\int d^3x\mbox{Tr}\left\{-\frac{1}{4}F_{\mu\nu}F^{\mu\nu}
+D_\mu \xi^\dagger D^\mu \xi
+i{\bar\Psi} D_\mu\Gamma^\mu\Psi\right.
\nonumber\\
&&\left.-g\left[{\bar\Psi}\Lambda\xi
+\xi^\dagger{\bar\Lambda}\Psi\right]
+\frac{i}{2}{\bar\Lambda}\Gamma^\mu D_\mu \Lambda
+\frac{\kappa}{2}\epsilon^{\mu\nu\lambda}
  \left[A_{\mu}\partial_{\nu}A_{\lambda}
           +\frac{2i}{3}gA_\mu A_\nu A_\lambda \right]
+\kappa\bar{\Lambda}\Lambda
\right\}.
\eea
The adjoint fields are the gauge boson $A_\mu$ (gluons)
and a Majorana fermion $\Lambda$ (gluinos);
the fundamental fields are the Dirac fermion $\Psi$ (quarks)
and a complex scalar $\xi$ (squarks).
The CS coupling $\kappa$ induces a mass for the adjoint fields
without breaking the supersymmetry; this reduces formation
of the long strings characteristic of super Yang--Mills theory.
The covariant derivatives are
\be
D_\mu\Lambda=\partial_\mu\Lambda+ig[A_\mu,\Lambda]\,,\;\;
D_\mu\xi=\partial_\mu\xi+igA_\mu\xi,  \;\;
D_\mu\Psi=\partial_\mu\Psi+igA_\mu\Psi.
\ee
The supersymmetry transformations are
\be
\delta A_\mu=\frac{i}{2}{\bar\varepsilon}\Gamma_\mu\Lambda,\;\;
\delta\Lambda=\frac{1}{4}F_{\mu\nu}\Gamma^{\mu\nu}\varepsilon,\;\;
\delta\xi=\frac{i}{2}{\bar\varepsilon}\Psi,\;\;
\delta\Psi=-\frac{1}{2}\Gamma^\mu\varepsilon D_\mu\xi.
\ee
We reduce this theory to 1+1 dimensions by taking the fields to
be independent of the transverse coordinate $x^2$.

As usual, there are constraints and not all of the fields are dynamical.  
To separate the dynamical fields, we first introduce components
for the Fermi fields and the supercharge as
\be
\Lambda=\left(\lambda,{\tilde\lambda}\right)^T\,,\qquad
\Psi=\left(\psi,{\tilde\psi}\right)^T\,,\qquad
Q=\left(Q^+,Q^-\right)^T.
\ee
Then, in light-cone gauge ($A^+=0$), the constraints are
\bea
\partial_-{\tilde\lambda}&=&-\frac{ig}{\sqrt{2}}
\left([A^2,\lambda]+i\xi\psi^\dagger-i\psi\xi^\dagger\right), \;\;
\partial_-{\tilde\psi}=-\frac{ig}{\sqrt{2}}A^2\psi+
\frac{g}{\sqrt{2}}\lambda\xi -\kappa\lambda/\sqrt{2}, \label{fcurrent} \\
\partial^2_-A^-&=&g\left\{i[A^2,\partial_-A^2]+
\frac{1}{\sqrt{2}}\{\lambda,\lambda\}
+\kappa \partial_- A^2
-ih\partial_-\xi\xi^\dagger+
i\xi\partial_-\xi^\dagger+\sqrt{2}\psi\psi^\dagger\right\}.
\eea
When these constraints are used to eliminate the
nondynamical fields, the supercharge becomes
\bea
Q^-&=&g\int dx^-\left\{
    2^{3/4}\left({\rm i}[A^2,\partial_-A^2]-\kappa\partial_-A^2
  +\frac{1}{\sqrt{2}}\{\lambda,\lambda\}\right)\frac{1}{\partial_-}\lambda 
  \right.  \\
&&-\frac{1}{\sqrt{2}}\left(i\sqrt{2}\xi\partial_
-\xi^\dagger-i\sqrt{2}\partial_-\xi\xi^\dagger+2\psi\psi^\dagger\right)
\frac{1}{\partial_-}\lambda 
\left.-2\left(\xi^\dagger A^2\psi+\psi^\dagger A^2\xi\right)\right\}.
\nonumber
\eea
The mode expansions of the dynamical fields are
\bea
\label{expandA2}
A^2_{ij}(0,x^-)&=&\frac{1}{\sqrt{4\pi}}\sum_{k=1}^{\infty}\frac{1}{\sqrt{k}}
\left(a_{ij}(k)e^{-ik\pi x^-/L}+a^\dagger_{ji}(k)e^{ik\pi x^-/L}\right),\\
\label{expandLambda}
\lambda_{ij}(0,x^-)&=&\frac{1}{2^{\frac{1}{4}}\sqrt{2L}}\sum_{k=1}^{\infty}
\left(b_{ij}(k)e^{-ik\pi x^-/L}+b^\dagger_{ji}(k)e^{ik\pi x^-/L}\right),\\
\label{expandxi}
\xi_i(0,x^-)&=&\frac{1}{\sqrt{4\pi}}\sum_{k=1}^{\infty}\frac{1}{\sqrt{k}}
\left(c_i(k)e^{-ik\pi x^-/L}+{\tilde c}^\dagger_{i}(k)e^{ik\pi
x^-/L}\right),\\
\label{expandPsi}
\psi_{i}(0,x^-)&=&\frac{1}{2^{\frac{1}{4}}\sqrt{2L}}\sum_{k=1}^{\infty}
\left(d_{i}(k)e^{-ik\pi x^-/L}+{\tilde d}^\dagger_{i}(k)e^{ik\pi
x^-/L}\right).
\eea 
The creation and annihilation operators obey the
commutation relations (for finite $N_c$)
\be
\left[a_{ij},a^\dagger_{kl}\right]=
\left(\delta_{il}\delta_{kj}-\frac{1}{N_c}\delta_{ij}\delta_{kl}\right),
\quad
\left\{b_{ij},b^\dagger_{kl}\right\}=
\left(\delta_{il}\delta_{kj}-\frac{1}{N_c}\delta_{ij}\delta_{kl}\right),
\ee
\be
\left[c_{i},c^\dagger_{j}\right]=\delta_{ij}\,,\quad
\left[{\tilde c}_{i},{\tilde c}^\dagger_{j}\right]=\delta_{ij},\quad
\left\{d_{i},{d}^\dagger_{j}\right\}=\delta_{ij} \quad
\left\{{\tilde d}_{i},{\tilde d}^\dagger_{j}\right\}=\delta_{ij}.
\ee

The solutions that we obtain are meson-like states
${\bar f}^\dagger_{i_1}(k_1) a^\dagger_{i_1i_2}(k_2)\dots
    b^\dagger_{i_ni_{n+1}}(k_{n-1})\dots f^\dagger_{i_p}(k_n)|0\rangle$,
where $f^\dagger=c^\dagger$ or $d^\dagger$, and glueball states
${\rm Tr}[a^\dagger_{i_1i_2}(k_1)\dots b^\dagger_{i_ni_{n+1}}(k_{n})]|0\rangle$.  
Because of the supersymmetry, either could be a boson or a fermion.
Because we work in the large-$N_c$ limit, there is no mixing
between these states, and they are composed of single traces.  This
simplifies the calculation, particularly with respect to the
size of the matrices that are diagonalized; however, study of
baryons will require finite $N_c$ or additional approximations.
To help reduce the size of the calculation further,
there is an additional $Z_2$ symmetry~\cite{Kutasov}
$a_{ij}(k,n^\perp)\rightarrow -a_{ji}(k,n^\perp)$,
$b_{ij}(k,n^\perp)\rightarrow -b_{ji}(k,n^\perp)$.
This divides the states between those with even and odd
numbers of gluons, and we diagonalize in each sector
separately.
A collection of results can be found in Ref.~\cite{SQCD-CS}.
The spectrum for mesons shows
the existence of light states at strong coupling,
a feature found previously in other supersymmetric
theories~\cite{LightStates}.

\section{CONCLUSION}

Methods for the nonperturbative solution of multidimensional
field theories are now available; they do, however, require
more development.  A number of calculations will soon be
undertaken to further demonstrate and improve this capability.
In Yukawa theory, the two-fermion sector is of interest,
particularly for a pseudo-scalar coupling with which the
deuteron might be modelled.  Quantum electrodynamics is
immediately treatable with the same techniques; the anomalous
magnetic moment of the electron can be computed.  The full
(2+1)-dimensional SQCD-CS theory can be solved, including
finite-$N_c$ corrections to allow baryons and meson-glueball
mixing.  This work should bring us closer to the goal of
being able to solve for hadron properties in quantum chromodynamics.

\vskip1cm
\noindent

\section*{ACKNOWLEDGMENTS}
The work summarized here was done in collaboration with
S.J. Brodsky and G. McCartor, and with 
S.S. Pinsky and U. Trittmann, and was 
supported in part by the US Department of Energy
and the Minnesota Supercomputing Institute.


\begin{thebibliography}{99}
%
\bibitem{DLCQreview} S.J.~Brodsky, H.-C.~Pauli, and S.S.~Pinsky,
Phys.\ Rep.\ \textbf{301} (1997) 299;
J.R. Hiller, Nucl.\ Phys.\ B (Proc.\ Suppl.) \textbf{90} (2000) 170.
%
\bibitem{Dirac} P.A.M. Dirac, 
Rev.\ Mod.\ Phys.\ \textbf{21} (1949) 392.
%
\bibitem{TwoParticles} S.J.~Brodsky, J.R.~Hiller, and G.~McCartor,
Ann.\ Phys.\ \textbf{305} (2003) 266 [hep-th/0209028].
%
\bibitem{SQCD-CS} J.R.~Hiller, S.S. Pinsky, and U.~Trittmann,
Phys.\ Rev.\ D \textbf{67} (2003) 115005 [hep-ph/0304147];
Nucl.\ Phys.\ B \textbf{661} (2003) 99 [hep-ph/0302119].
For earlier work on fundamental matter in SDLCQ, see
O. Lunin and S. Pinsky, Phys.\ Rev.\ D \textbf{63} (2001) 045019.
%
\bibitem{PauliVillars} W. Pauli and F. Villars,
   Rev.\ Mod.\ Phys.\ \textbf{21} (1949) 4334.
%
\bibitem{bhm} S.J.~Brodsky, J.R.~Hiller, and G.~McCartor,
Phys.\ Rev.\ D \textbf{58} (1998) 025005 [hep-th/9802120];
Phys.\ Rev.\ D \textbf{60} (1999) 054506 [hep-ph/9903388];
Ann.\ Phys.\ \textbf{296} (2002) 406 [hep-th/0107246].
%
\bibitem{Special} S.J.~Brodsky, J.R.~Hiller, and G.~McCartor,
Phys.\ Rev.\ D \textbf{64} (2001) 114023 [hep-ph/0107038].
%
\bibitem{SDLCQreview} O.~Lunin and S.~Pinsky,
   in {\em New Directions in Quantum Chromodynamics}, 
   edited by C.-R.~Ji and D.-P.~Min,
   AIP Conf.\ Proc.\ No.\ 494 (AIP, Melville, NY, 1999), p.~140,
   [hep-th/9910222].
%
\bibitem{TransLattice} M.~Burkardt and S.~Dalley,
Prog.\ Part.\ Nucl.\ Phys. \ \textbf{48} (2002) 317 [hep-ph/0112007]; 
S.~Dalley and B.~van~de~Sande, 
Phys.\ Rev.\ D \textbf{67} (2003) 114507 [hep-ph/0212086].
%
\bibitem{Glazek}  St.D.~Glazek and J.~Mlynik,
Phys.\ Rev.\ D \textbf{67} (2003) 045001 [hep-th/0210110];
St.D.~Glazek and M.~Wieckowski,
Phys.\ Rev.\ D \textbf{66} (2002) 016001 [hep-th/0204171].
%
\bibitem{PauliBrodsky} H.-C.~Pauli and S.J.~Brodsky,
Phys.\ Rev.\ D \textbf{32} (1985) 1993; \textbf{32} (1985) 2001.
For earlier, more formal work, see
T. Maskawa and K. Yamawaki, Prog.\ Theor.\ Phys.\ \textbf{56} (1976) 270.
%
\bibitem{Sakai} Y.~Matsumura, N.~Sakai, and T.~Sakai,
Phys.\ Rev.\ D \textbf{52} (1995) 2446 [hep-th/9504150].
%
\bibitem{Lanczos} C. Lanczos,
   J. Res.\ Nat.\ Bur.\ Stand.\ \textbf{45} (1950) 255;
   J. Cullum and R.A. Willoughby,
   {\em Lanczos Algorithms for Large Symmetric Eigenvalue Computations}
   (Birkhauser, Boston, 1985), Vol. I and II.
%
\bibitem{biorthogonal} J.H. Wilkinson,
   {\em The Algebraic Eigenvalue Problem} (Clarendon, Oxford, 1965);
   Y. Saad, Comput.\ Phys.\ Commun.\ \textbf{53} (1989) 71;
   S.K. Kin and A.T. Chronopoulos,
   J. Comp.\ and Appl.\ Math \textbf{42} (1992) 357.
%
\bibitem{Cullum} J. Cullum and R.A. Willoughby,
   in {\em Large-Scale Eigenvalue Problems},
   edited by J. Cullum and R.A. Willoughby,
   Math.\ Stud.\ \textbf{127}
   (Elsevier, Amsterdam, 1986), p.~193.
%
\bibitem{Kutasov} D.~Kutasov, Nucl.\ Phys.\ B \textbf{414} (1994) 33.
%
\bibitem{LightStates} J.R.~Hiller, S.S.~Pinsky, and U.~Trittmann,
Phys.\ Rev.\ Lett.\  \textbf{89} (2002) 181602 [hep-th/0203162];
J.R.~Hiller, S.S.~Pinsky, and U.~Trittmann,
Phys.\ Lett.\ B \textbf{541} (2002) 396 [hep-th/0206197].
%
\end{thebibliography}
\end{document}